\documentclass[trackchanges,twocolumn]{aastex701}
\usepackage{amsmath}
\usepackage{booktabs} 
\usepackage{xcolor}

\begin{document}

\title{Galaxy--LRD Strong Lenses: A Missing Population?}

\correspondingauthor{Zizhao He}

\author[0000-0001-8554-9163]{Zizhao He}
\affiliation{Department of Physics, Nanchang University, Nanchang 330031, China}
\affiliation{Center for Relativistic Astrophysics and High Energy Physics, Nanchang University, Nanchang 330031, China}
\affiliation{Purple Mountain Observatory, Chinese Academy of Sciences, Nanjing, Jiangsu 210023, China}
\email[show]{zzhe@ncu.edu.cn}

\author[0000-0001-6800-7389]{Nan Li}
\affiliation{National Astronomical Observatories, Chinese Academy of Sciences, Beijing 100101, China}
\email{nan.li@nao.cas.cn}

\author[0000-0002-1318-8343]{Simon Dye}
\affiliation{School of Physics and Astronomy, University of Nottingham, Nottingham NG7 2RD, UK}
\email{simon.dye@nottingham.ac.uk}

\author[0000-0002-8700-3671]{Xinzhong Er}
\affiliation{Tianjin Astrophysics Center, Tianjin Normal University, Tianjin 300387, China}
\email{phioen@163.com}

\author[0000-0002-0752-6457]{Fuwen Shu}
\affiliation{Department of Physics, Nanchang University, Nanchang 330031, China}
\affiliation{Center for Relativistic Astrophysics and High Energy Physics, Nanchang University, Nanchang 330031, China}
\email{shufuwen@ncu.edu.cn}

\begin{abstract}
    The physical nature of Little Red Dots (LRDs) remains uncertain, although these abundant, compact, and red sources may offer important insights into early black-hole growth and galaxy formation. Strong gravitational lensing can magnify LRDs and spatially resolve their internal structure, thereby helping to discriminate among competing physical scenarios. However, no galaxy-scale strongly lensed LRD has yet been securely confirmed. To predict the abundance of such systems in current and future surveys and to guide dedicated searches, we present the first benchmark estimate of the detectable population of galaxy-scale lensed LRDs by combining literature-based LRD source models with a population of foreground early-type-galaxy deflectors. Our Monte Carlo simulation spans $50~{\rm deg}^{2}$ and contains 270,713 LRDs and 5,460,841 deflectors. We predict idealized surface densities of $10.70\pm3.76~{\rm deg}^{-2}$ for doubles and $0.64\pm0.69~{\rm deg}^{-2}$ for quads. After accounting for the JWST point-spread function and survey limiting magnitudes, the detectable surface densities decrease to $3.70\pm1.89~{\rm deg}^{-2}$ and $0.52\pm0.58~{\rm deg}^{-2}$, respectively. For the de-duplicated $0.66~{\rm deg}^{2}$ footprint covered by COSMOS-Web, PRIMER-UDS, PRIMER-COSMOS, CEERS, JADES GOODS-S, and JADES GOODS-N, for which the reported limiting depths are combined through area-weighted averaging in flux space, the predicted probabilities of detecting no systems are $8.6\%$ for doubles and $70.8\%$ for quads.
\end{abstract}



\section{Introduction}
    Little Red Dots (LRDs) are a recently identified population of compact, red, high-redshift sources revealed by deep imaging with the James Webb Space Telescope \citep{Matthee2024,Kokorev2024,Kocevski2025,Akins2025}. Growing spectroscopic evidence indicates that many LRDs, particularly those exhibiting broad Balmer emission lines, host actively accreting black holes \citep{Matthee2024,Kocevski2025,Furtak2024}. The central question is therefore no longer simply whether an active galactic nucleus (AGN) is present, but rather what powers the compact rest-frame optical emission and how much each physical component contributes. The observed radiation may originate directly from black-hole accretion, from the absorption, scattering, and reprocessing of AGN emission by dense surrounding gas, from an extremely compact stellar component, or from some combination of these processes \citep{Killi2024,Baggen2024,Chen2025}.

    Strong gravitational lensing offers several complementary routes for distinguishing among these physical scenarios \citep[see, e.g.,][and references therein]{Treu2010,Shajib2024,Natarajan2024}. By magnifying the background source and stretching it over a larger angular area, lensing can help separate the extremely compact central emission from the more extended host galaxy. This enables more robust measurements of the host morphology, size, stellar mass, and spatial relationship to the nucleus \citep{Furtak2024}. The increased observed flux can also improve the signal-to-noise ratio of spectroscopic observations, making intrinsically weak emission features accessible and tightening constraints on the density, obscuration, and physical conditions of the gas surrounding the black hole \citep{Furtak2024,Juodzbalis2026}.

    A2744-QSO1 provides a clear proof of concept for the scientific value of strongly lensed LRDs. This source at $z=7.045$ is observed as three images behind the Abell 2744 galaxy cluster. The lensing-enhanced flux observed with JWST has enabled the identification of a heavily reddened broad-line AGN, measurements of its host galaxy, and a direct dynamical estimate of a central black-hole mass of approximately $5\times10^7 M_\odot$ \citep{Furtak2024,Juodzbalis2026}. The multiple images and their time delays have also enabled independent variability tests of the AGN interpretation \citep{Furtak2025}. 

    Galaxy-scale lensed LRDs would complement cluster-lensed systems by offering several distinct advantages. Although individual galaxy clusters have larger strong-lensing cross-sections and can produce extreme magnifications, massive clusters are rare, whereas foreground galaxies are vastly more numerous. Galaxy-scale lenses may therefore produce a larger and more statistically useful population of lensed LRDs. Their mass distributions are also typically simpler and more compact than those of cluster lenses. Consequently, galaxy-scale systems can in principle provide more precise magnification estimates and more robust reconstructions of the intrinsic morphology, size, luminosity, and black-hole-to-galaxy mass ratio of the background LRD. This is analogous to time-delay cosmography, for which galaxy-scale lensed quasars are generally preferred because their lens potentials can be constrained more accurately than those of cluster-scale systems \citep[e.g.][]{Wong2020,Shajib2024,Tdcosmo2025}. 

    No galaxy-scale strongly lensed LRD has yet been securely confirmed, limiting the use of strong lensing as a high-resolution probe of the LRD. We therefore present the first benchmark estimate of the expected abundance of galaxy-scale lensed LRDs in current and future surveys, primarily to assess whether detectable systems should already exist in available JWST data and to guide future dedicated searches. Following the forward-modeling approach used in strong-lens population forecasts, including the OM10 \citep{OguriMarshall2010} and the studies of \citet{Collett2015} and \citet{Cao2024}, we combine literature-based models of the LRD population \citep{Kokorev2024} with a foreground population of early-type galaxy lenses constructed from observed velocity-dispersion and structural distributions \citep{Choi2007,Hyde2009,Collett2015}. We then simulate the resulting galaxy-scale strong-lensing systems and apply an observational transfer function that accounts for two dominant selection effects: the finite angular resolution imposed by the PSF and the limiting magnitudes of JWST imaging \citep{Baggen2024,Casey2023,Donnan2024,McLeod2024}. This framework predicts both the intrinsic strong-lensing probability of the LRD population and the fraction of systems that remain detectable under realistic observing conditions.

    This paper is organized as follows. Section~\ref{sec:method} describes the construction of the background LRD and foreground ETG populations, the simulation of ideal strong-lensing systems, and the observational transfer function. Section~\ref{sec:res} presents the predicted properties and surface densities of observable lensed LRDs and quantifies the significance of a null detection as a function of survey area. Section~\ref{sec:diss} discusses the impact of foreground lens-galaxy light on the detectability of lensed LRDs. Section~\ref{sec:conclu} summarizes our main conclusions.

\section{Methodology}
\label{sec:method}
    Our methodology consists of four main steps. First, we construct an area-based mock catalog of background LRDs from a parameterized rest-frame UV luminosity function, assigning each source a redshift, intrinsic UV luminosity, and synthetic JWST/NIRCam photometry. Second, we generate a foreground population of early-type galaxies (ETGs) with redshifts, velocity dispersions, projected shapes, sizes, and broad-band fluxes. Third, we randomly pair the foreground deflectors with the background LRDs and solve the lens equation to identify multiple-image systems, thereby defining the intrinsic, or ideal, strongly lensed population. Finally, we apply an observational transfer function based on the angular resolution and limiting depths of representative JWST/NIRCam imaging to determine which simulated systems would be detectable in existing surveys. The following subsections describe these components in turn.

\subsection{Background LRD population}
\label{sec:source}

    \begin{figure*}
        \centering
        \includegraphics[scale=0.45]{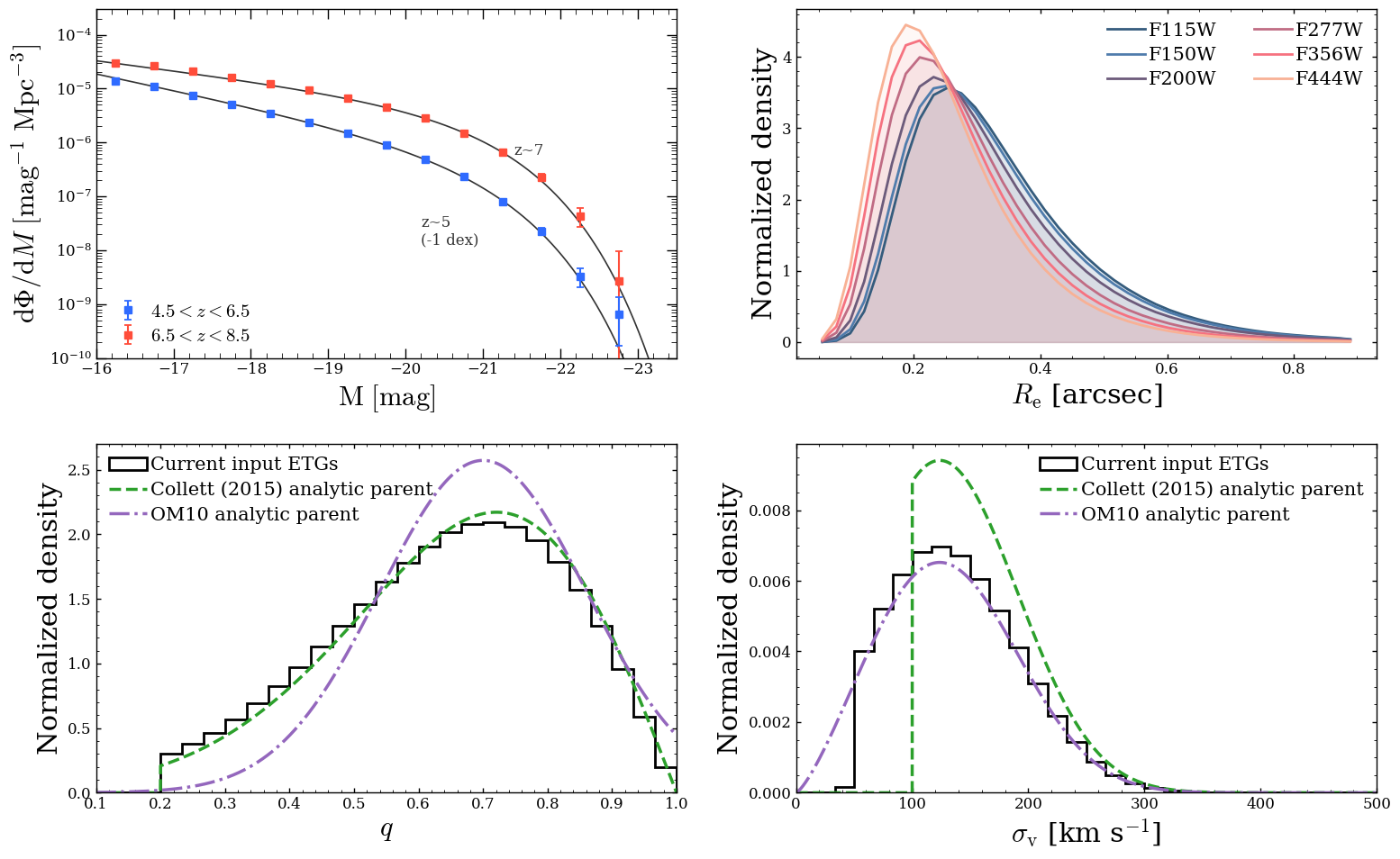}
        \caption{Properties of the mock LRD and foreground-ETG populations. Upper-left panel: rest-frame UV luminosity functions of the mock LRD catalog in the two adopted redshift bins. The black curves show the input K24 Schechter functions, and the points show the corresponding distributions recovered from the mock catalog; the $z\sim5$ luminosity function is shifted downward by 1 dex for clarity. Upper-right panel: normalized distributions of the effective radius of ETGs in six simulated bands. Bottom-left and bottom-right panels: normalized distributions of the projected axis ratio, $q_{\rm lens}$, and velocity dispersion, $\sigma_{\rm v}$, respectively, for the mock foreground ETGs, compared with the analytic parent distributions adopted by OM10 and \citet{Collett2015}.}
    
        \label{fig:lrd}
    \end{figure*}
    We adopt the rest-frame UV luminosity function (UVLF) measured by
    \citet[][K24 hereafter]{Kokorev2024} as our fixed fiducial model because it is particularly well suited to this study. K24 provides one of the few explicit Schechter parameterizations of the LRD UVLF at rest-frame $1450\,\text{\AA}$, with separate parameters for $4.5<z<6.5$ and $6.5<z<8.5$; we therefore adopt these intervals as the redshift range of our mock LRD population.
    The measurement is also based on approximately $640\,\mathrm{arcmin}^{2}$ distributed across four independent, unlensed JWST blank fields, providing a comparatively broad and field-diverse estimate of the LRD abundance.
    
    Furthermore, the imaging data were processed homogeneously, and candidates were identified across all fields using a common set of photometric color and compactness criteria. The resulting number densities were corrected object by object using the $1/V_{\max}$ method.
    We use this parameterized UVLF to generate an area-based mock catalog of intrinsic and observed LRD properties.
    
    Following K24, we model the redshift dependence of the UVLF using their two-bin Schechter parameterization \citep{Schechter1976}. The differential UVLF is
    \begin{equation}
    \begin{aligned}
    \phi(M,z)
    &=0.4\ln 10\,\phi_\ast(z)\,
    t(M,z)^{\alpha(z)+1}
    e^{-t(M,z)},\\
    t(M,z)
    &\equiv 10^{0.4[M_\ast(z)-M]}.
    \end{aligned}
    \end{equation}
    We adopt the best-fitting Schechter parameters reported by K24 and keep them fixed within each redshift bin. We do not propagate the uncertainties in $(M_\ast,\phi_\ast,\alpha)$ because their joint posterior distributions and covariance matrices are not publicly available. Observational uncertainties are nevertheless included separately, as described later in this section.
    
    The bin-dependent best-fitting parameters are
    \begin{equation}
    (M_\ast,\phi_\ast,\alpha)
    =
    (-20.64,\,0.008\times10^{-3},\,-1.76)
    \end{equation}
    for $4.5<z<6.5$, and
    \begin{equation}
    (M_\ast,\phi_\ast,\alpha)
    =
    (-20.67,\,0.005\times10^{-3},\,-1.46)
    \end{equation}
    for $6.5<z<8.5$.

    The UVLF determines both the comoving source density and the source-redshift distribution. For each redshift bin, we integrate over the adopted magnitude range to obtain the comoving number density,
    \begin{equation}
    n_{\rm src}(z)
    =
    \int_{-23.5}^{-16.0}
    \phi(M,z)\,dM.
    \end{equation}
    Because the Schechter parameters are fixed within each K24 bin, $n_{\rm src}(z)$ is piecewise constant within each bin. We then draw source redshifts from the comoving-volume-weighted distribution
    \begin{equation}
    p(z)\propto
    n_{\rm src}(z)
    \frac{dV_{\rm c}}{dz\,d\Omega},
    \end{equation}
    where $dV_{\rm c}/dz\,d\Omega$ is the differential comoving volume per unit redshift per unit solid angle. After assigning a source redshift, we sample its intrinsic $M_{1450}$ from the corresponding bin-specific Schechter distribution.

    For a survey area $\Omega$, the expected number of sources is
    \begin{equation}
    N_{\rm exp}^{\rm src}
    =
    \Omega
    \int_{4.5}^{8.5}
    n_{\rm src}(z)
    \frac{dV_{\rm c}}{dz\,d\Omega}
    \,dz.
    \end{equation}
    We draw the realized catalog size from a Poisson distribution with mean $N_{\rm exp}^{\rm src}$.

    We perturb the intrinsic redshift and magnitude to mimic measurement uncertainties and obtain the observed quantities $(z_{\rm obs},M_{1450,{\rm obs}})$. The redshift perturbation is drawn from a truncated Gaussian with $\sigma_z=f_z(1+z)$, for which we adopt $f_z=0.02$, and the magnitude perturbation is drawn from a Gaussian with $\sigma_M=0.11$ mag. Both values are motivated by the K24 LRD catalog: the median normalized photometric-redshift uncertainty, $0.5(z_{\rm phot,84}-z_{\rm phot,16})/(1+z_{\rm phot})$, is approximately $0.02$, and the median uncertainty in $M_{\rm UV}$, reported as ${\tt muv_{err}}$, is approximately $0.11$ mag.

    The mock UVLFs closely reproduce the input K24 Schechter functions in both redshift bins, confirming that the sampling procedure recovers the adopted source population. The resulting UVLFs are shown in the upper-left panel of Figure~\ref{fig:lrd}. For clarity, the luminosity function in the lower-redshift bin is shifted downward by 1 dex.

    We optionally synthesize JWST broad-band AB magnitudes for each mock LRD using the spectroscopically confirmed LRD spectra from K24 as empirical templates. Of the 15 spectroscopically confirmed LRDs reported by K24, we select four objects with complete wavelength coverage across the relevant bands and high spectral quality, requiring an average signal-to-noise ratio greater than 8. The selected templates have redshifts of $z=4.95$, 5.28, 6.69, and 6.99.

    Each mock LRD is assigned a template that provides finite synthetic photometry in all required NIRCam bands. If multiple templates satisfy this criterion, we select the one closest in redshift to the mock source. We then rescale the chosen template to match the simulated $M_{1450}$ at the target redshift and compute synthetic photometry in the six NIRCam bands \texttt{F115W}, \texttt{F150W}, \texttt{F200W}, \texttt{F277W}, \texttt{F356W}, and \texttt{F444W}.

\subsection{Foreground ETG population}

    We generate the ETG population with the \texttt{DeflectorPopulation} module of
    \texttt{SimCsstLens}\footnote{\url{https://github.com/caoxiaoyue/sim\_csst\_lens}}
    \citep{Cao2024}. We adopt the velocity-dispersion range
    \begin{equation}
    50 \leq \sigma_{\rm v}/({\rm km\,s^{-1}}) \leq 400,
    \end{equation}
    and restrict the parent lens-redshift range to
    \begin{equation}
    z_{\rm lens} \le 4.5.
    \end{equation}

    The module assigns each mock ETG a lens redshift, projected axis ratio, velocity dispersion, effective radius, and absolute $r$-band magnitude; further details are given by \citet{Cao2024}. For a sky area $\Omega$, the code first computes the corresponding sky fraction,
    \begin{equation}
    f_{\rm sky} = \Omega / 41252.96~{\rm deg}^2,
    \end{equation}
    and then evaluates the expected number of ETGs by integrating the deflector population over the corresponding comoving volume. Specifically, the model assumes that the comoving velocity-dispersion function does not evolve with redshift, so that
    \begin{equation}
    \frac{{\rm d}N_{\rm ETG}}{{\rm d}z\,{\rm d}\Omega}
    =
    \left[
    \int_{\sigma_{\rm v,min}}^{\sigma_{\rm v,max}}
    \phi(\sigma_{\rm v})\,{\rm d}\sigma_{\rm v}
    \right]
    \frac{{\rm d}V_{\rm c}}{{\rm d}z\,{\rm d}\Omega}.
    \end{equation}

    Lens redshifts are drawn over the adopted redshift range from the normalized distribution defined by this differential number density, which is proportional to the differential comoving-volume element ${\rm d}V_{\rm c}/({\rm d}z,{\rm d}\Omega)$. For each ETG, the projected axis ratio, $q_{\rm lens}$, is drawn from the Rayleigh distribution adopted by \citet{Collett2015}. The velocity dispersion, $\sigma_{\rm v}$, is sampled from the velocity-dispersion function of \citet{Choi2007} and is then used to assign the effective radius, $R_{\rm e}$, and absolute $r$-band magnitude through the $r$-band fundamental-plane relation of \citet{Hyde2009}.
    
    We augment each ETG with approximate JWST photometry. Using an SWIRE elliptical-galaxy template and adopted $5\sigma$ depth parameters in the six NIRCam bands
    \texttt{F115W}, \texttt{F150W}, \texttt{F200W}, \texttt{F277W}, \texttt{F356W}, and \texttt{F444W},
    we generate mock foreground magnitudes and their associated uncertainties.

    We account for the wavelength dependence of the apparent lens-galaxy size by propagating the effective radius into each JWST band with a power-law size--wavelength correction,
    \begin{equation}
    R_{\rm e}(\lambda_{\rm rest}) = R_{\rm e}(\lambda_{0,{\rm rest}})
    \left(\frac{\lambda_{\rm rest}}{\lambda_{0,{\rm rest}}}\right)^{\beta}.
    \end{equation}
    We adopt $\lambda_{0,{\rm rest}}=0.62\,\mu{\rm m}$, approximately the effective wavelength of the SDSS $r$ band \citep{Fukugita1996,Doi2010}, and $\beta=-0.25$, corresponding to the mean size gradient $\Delta \log R_{\rm eff}/\Delta \log \lambda$ measured for early-type galaxies by \citet{vanderWel2014}.

    Figure~\ref{fig:lrd} provides a joint summary of the input LRD and foreground-ETG populations. The upper-left panel shows the mock LRD UVLFs discussed above, while the upper-right panel shows the normalized distributions of the effective radii of the foreground ETGs in the six simulated bands. The bottom-left and bottom-right panels show the projected axis-ratio distribution, $q_{\rm lens}$, and the velocity-dispersion distribution, $\sigma_v$, respectively, for the foreground ETGs. For comparison, we show the analytic parent distributions adopted in OM10 and \citet{Collett2015}, thereby avoiding differences introduced by study-specific selection criteria. The $q_{\rm lens}$ and $\sigma_{\rm v}$ distributions in our mock sample agree well with those adopted in both studies.

\subsection{Simulating the ideal lenses}

    We simulate lens--source pairings by randomly sampling background sources on the source plane around each ETG. Because all ETGs have $z_{\rm d}<4.5$, whereas the LRD catalog is restricted to $z_{\rm s}\geq4.5$, no additional lens--source redshift cut is required.

    We set the source-plane sampling area so that each lens contains one source on average. The source surface density is estimated directly from the mock LRD catalog as
    \[
    \Sigma_{\rm src}
    =
    \frac{N_{\rm src,cat}}{A_{\rm cat}}
    =
    4.1777\times10^{-4}\ {\rm arcsec}^{-2},
    \]
    where \(N_{\rm src,cat}=270713\) is the number of catalog sources and \(A_{\rm cat}=50~{\rm deg}^2\) is the catalog area. For each foreground lens, we define a square source-plane sampling region with area
    \[
    A_{\rm box}=\Sigma_{\rm src}^{-1},
    \]
    and define \(L_{\rm box}\equiv\sqrt{A_{\rm box}}\) as the side length of this region. This gives
    \[
    A_{\rm box}\approx 2393.68~{\rm arcsec}^2,
    \qquad
    L_{\rm box}\approx 48.93''.
    \]

    The number and positions of candidate sources associated with each lens are sampled stochastically. We draw
    \[
    N_{\rm src}\sim {\rm Poisson}(\Sigma_{\rm src}A_{\rm box})
    ={\rm Poisson}(1),
    \]
    for which
    \[
    \begin{cases}
    P(N_{\rm src}=0)=P(N_{\rm src}=1)=e^{-1}=36.79\%,\\
    P(N_{\rm src}\geq 2)=1-2e^{-1}=26.42\%.
    \end{cases}
    \]
    The source positions are sampled uniformly within the box,
    \[
    x_{\rm src},\,y_{\rm src}\sim
    {\rm Uniform}\left(-\frac{L_{\rm box}}{2},\,\frac{L_{\rm box}}{2}\right),
    \]
    and each source is randomly assigned the redshift and intrinsic photometric properties of an object from the mock LRD catalog.

    To retain field-by-field information in the simulated catalog, we divide the full $50~{\rm deg}^2$ realization into 50 bookkeeping regions, each representing $1~{\rm deg}^2$. After randomly sampling the foreground ETG catalog, we assign each lens a sequential index $i$ and define
    \[
    {\tt square\_degree\_id}
    =
    \left\lfloor
    \frac{i-1}{N_{\rm ETG}/A_{\rm sim}}
    \right\rfloor+1,
    \]
    where \(N_{\rm ETG}\) is the number of sampled foreground galaxies and \(A_{\rm sim}=50~{\rm deg}^2\). The resulting identifier ranges from 1 to 50 and is stored in every simulated strong-lens record; systems associated with the same foreground lens therefore inherit the same identifier. This identifier is a survey-area bookkeeping label rather than a simulated sky coordinate. It allows us to calculate lensing rates and their field-to-field scatter directly from the counts in the 50 individual square-degree regions (See Section~\ref{sec:res}).
    
    We model the mass distribution of each foreground ETG as a singular isothermal ellipsoid \citep[SIE,][]{Kormann1994}. Following OM10, we add an external-shear term to account for tidal perturbations from the lens environment and matter along the line of sight. For each simulated ETG, the external-shear amplitude, $\gamma_{\rm ext}$, and position angle, $\phi_{\rm ext}$, are drawn together and with replacement from the corresponding entries in the OM10 mock catalog\footnote{\url{https://github.com/drphilmarshall/OM10/blob/master/data/qso_mock.fits}.}. The same shear realization is used for all candidate sources associated with a given foreground ETG.
    The full lens equation is therefore solved using an SIE plus external-shear model.
    
    We reduce the computational cost by pre-selecting lens--source pairs before solving the full lens equation. For each pair, we compute $\theta_{\rm E}$ from $(z_{\rm d},z_{\rm s},\sigma_{\rm v})$, where $(x_{\rm s},y_{\rm s})$ denotes the source position relative to the center of the foreground lens in the source plane, and require
    \[
    \sqrt{x_{\rm s}^2+y_{\rm s}^2}<\theta_{\rm E}.
    \]
    Only pairs that satisfy this criterion are passed to the full lens-equation solver, and systems with $n_{\rm images}\geq2$ are retained as strong lenses.
    
    \subsection{The transfer function}
    \label{sec:transfer_function}

    We convert the intrinsic lensing rates into observable numbers with a simple transfer function that accounts for two dominant selection effects: the finite angular resolution set by the point-spread function (PSF) and the limiting magnitude of JWST/NIRCam imaging \citep[e.g.][]{Rieke2023,Rigby2023,Bagley2023}. We evaluate the detectability of each mock lensed LRD independently in the six NIRCam bands considered here.

    To define representative survey depths, we compile the reported $5\sigma$ AB limiting magnitudes for six major JWST/NIRCam extragalactic survey fields: COSMOS-Web, PRIMER-UDS, PRIMER-COSMOS, CEERS, JADES GOODS-S, and JADES GOODS-N \citep{Bagley2023,Casey2023,Donnan2024,McLeod2024}. Following K24, we adopt the $5\sigma$ AB limiting magnitudes measured within $0\farcs36$ apertures. The nominal footprint areas are 1789, 234, 144, 100, 67, and 58 ${\rm arcmin}^{2}$, respectively. To avoid double counting, the COSMOS-Web area adopted here excludes the region overlapping with PRIMER-COSMOS. Because AB magnitudes are logarithmic, we first convert each limiting magnitude to a limiting flux density,
    \begin{equation}
    f_{\nu,{\rm lim},b,j}
    =
    f_{\nu,0}\,10^{-0.4m_{{\rm lim},b,j}},
    \end{equation}
    and then calculate its area-weighted mean,
    \begin{equation}
    \overline{f}_{\nu,{\rm lim},b}
    =
    \frac{\displaystyle\sum_{j\in\mathcal{J}_b} A_j
    f_{\nu,{\rm lim},b,j}}
    {\displaystyle\sum_{j\in\mathcal{J}_b} A_j}.
    \end{equation}
    The representative limiting magnitude is obtained by converting the mean
    limiting flux density back to AB magnitude,
    \begin{equation}
    \overline{m}_{{\rm lim},b}
    =
    -2.5\log_{10}
    \left(
    \frac{\overline{f}_{\nu,{\rm lim},b}}{f_{\nu,0}}
    \right),
    \end{equation}
    where $A_j$ is the adopted area of field $j$, $m_{{\rm lim},b,j}$ is its reported $5\sigma$ depth in filter $b$, $f_{\nu,0}$ is the AB zero-point flux density, and $\mathcal{J}_b$ contains only fields with an available depth measurement in that filter. Thus, fields with missing measurements are omitted for that filter and the remaining area weights are renormalized. All six fields contribute in F115W, F150W, F277W, and F444W, corresponding to a total area of $2392~{\rm arcmin}^{2}$. COSMOS-Web does not provide the F200W and F356W measurements used in our compilation, so the weighted averages in these two filters are based on the remaining five fields, with a total contributing area of $603~{\rm arcmin}^{2}$. This procedure gives the representative limiting magnitudes
    \begin{equation}
    \label{eq:weighted_depth}
    m_{\rm lim} = (27.00,\,27.19,\,28.32,\,27.87,\,28.69,\,27.73)
    \end{equation}
    for F115W, F150W, F200W, F277W, F356W, and F444W, respectively. The corresponding PSF full widths at half maximum are
    \begin{equation}
    {\rm FWHM}_{\rm PSF} = (0.040,\,0.050,\,0.066,\,0.092,\,0.116,\,0.145)\arcsec.
    \end{equation}
    These values are based on the expected NIRCam PSF sizes in the corresponding filters \citep{Rieke2023} and are consistent with the performance reported in commissioning and empirical PSF studies \citep{Rigby2023,Zhuang2024}. Although the PSF varies among filters, we adopt a conservative resolution criterion based on the broadest PSF, requiring the relevant image separation to exceed $0\farcs145$, the F444W PSF FWHM. We use these limits to determine whether the multiple images are sufficiently resolved and bright to be identified as a lensed system \citep[e.g.][]{Collett2015}.

    A mock system is classified as observable if it satisfies both the PSF-resolution and flux criteria in at least one of the six JWST bands. For doubles, both images must be detectable. For quads, we require the third-brightest image to exceed the limiting magnitude, following the image-ranking convention of OM10. This conservative criterion reflects the practical consideration that a nominal quad with only two detectable images would likely be classified as a double or an ambiguous lens candidate rather than as a secure quad.

\section{Results}
\label{sec:res}

\begin{figure*}
    \includegraphics[scale=0.45]{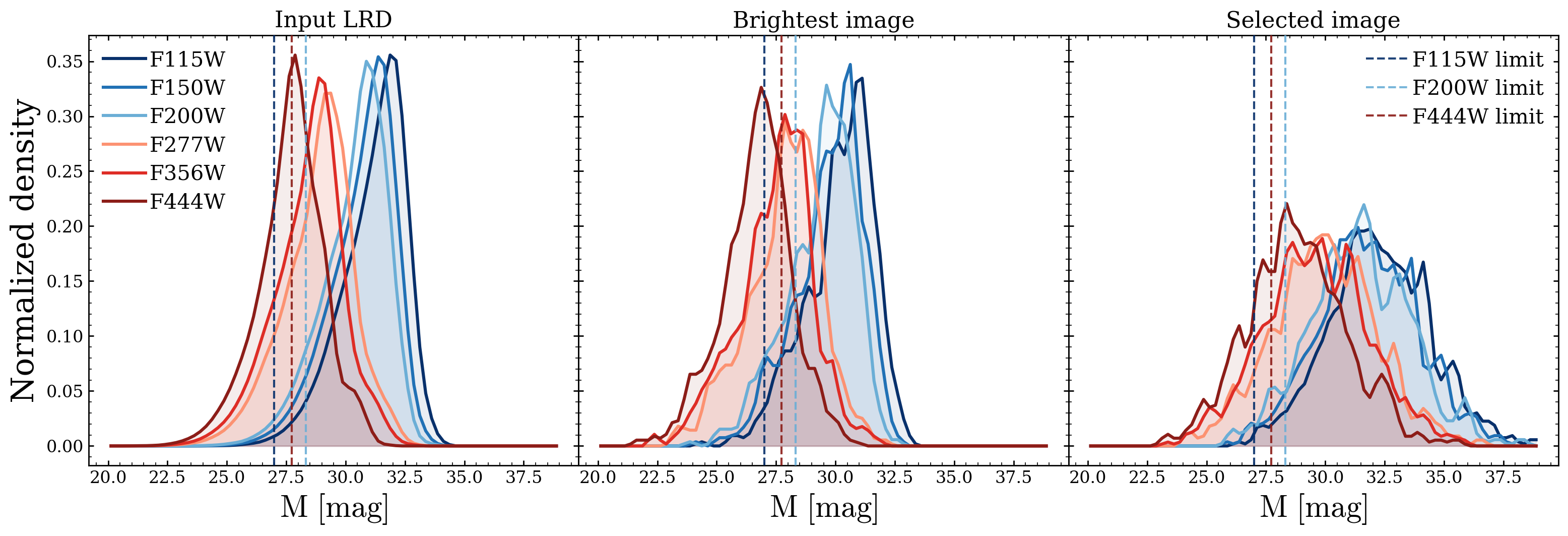}
    \includegraphics[scale=0.45]{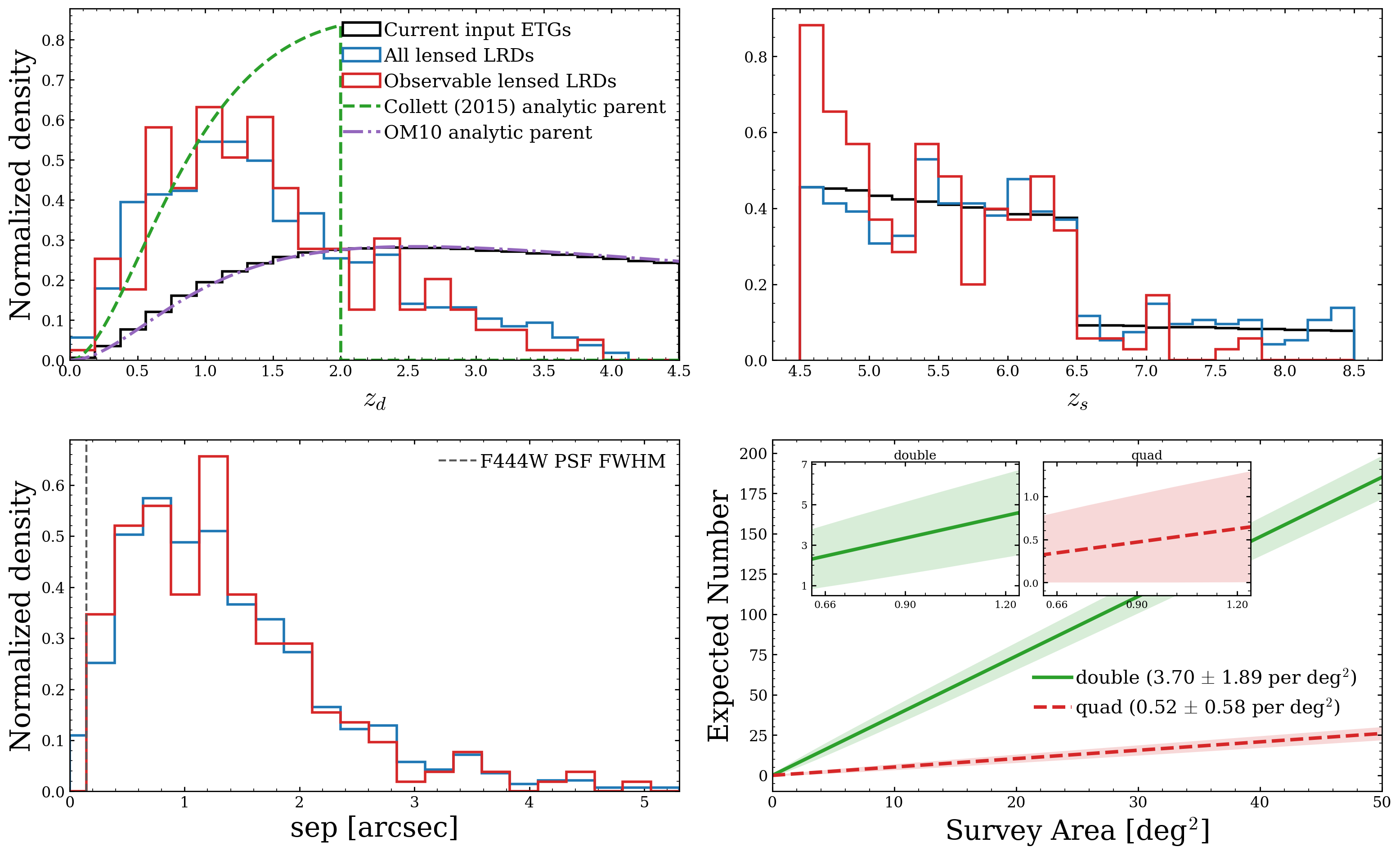}
    \caption{Top row: magnitude distributions of the input LRD catalog (left), the brightest lensed image in each multiple-image system (middle), and the selected lensed image (right), defined as the second-brightest image for doubles and the third-brightest image for quads. The vertical dotted lines indicate the adopted representative $5\sigma$ limiting magnitudes in the three NIRCam bands. Middle row: redshift distributions of mock lensed LRDs, showing the lens redshift (left) and source redshift (right). Bottom row: distribution of the maximum image separation for mock lensed LRDs (left) and expected number of observable lensed LRDs as a function of cumulative JWST survey area (right). The vertical dashed line in the separation panel marks the adopted F444W PSF FWHM. In the survey-area panel, the shaded bands show the expected field-to-field scatter inferred from the 50 simulated $1~{\rm deg}^{2}$ regions, and the two insets provide enlarged views of the double and quad predictions over $0.62$--$1.24~{\rm deg}^{2}$.}

    \label{fig:lensed_lrd}
\end{figure*}

    The simulation yields 567 ideal lensed LRDs and 211 observable systems from 5,460,841 ETGs and 270,713 LRDs over a $50~{\rm deg}^2$ survey area. Figure~\ref{fig:lensed_lrd} shows the relevant distributions. Among the ideal systems, one foreground ETG lenses two distinct LRD sources. However, this double-source system does not pass our observational selection criteria and is therefore not included in the observable sample. 

    Lensing brightens a fraction of the LRD population, but the image used for detection often remains close to the survey limits. The upper panels of Figure~\ref{fig:lensed_lrd} show the magnitude distributions of the input LRDs, the brightest lensed image in each multiple-image system, and the selected lensed image---the second-brightest image for doubles and the third-brightest image for quads---in the six NIRCam bands. The input LRD catalog is generally fainter than the adopted limiting magnitudes, especially in the bluer bands. Magnification shifts some sources to brighter apparent magnitudes, as shown by the brightest-image distributions, whereas the selected-image distributions remain broad and extend close to the limiting magnitudes.

    Observational selection changes the lens-redshift distribution only mildly. The full lensed sample spans $0.0487$--$4.0497$ in $z_d$, with a median of $1.3371$, whereas the observable sample spans $0.0918$--$3.8656$, with a median of $1.3476$ (middle-left panel of Figure~\ref{fig:lensed_lrd}). The observable sample therefore remains broadly distributed in lens redshift and closely follows the full lensed sample. For comparison, the redshift distribution of our input ETGs is nearly identical to the OM10 analytic parent distribution, because both are constructed from a redshift-independent comoving deflector abundance and extend across the full redshift interval shown here. The analytic parent distribution adopted by \citet{Collett2015}, by contrast, is restricted to $z_d\leq2$ and consequently has a sharp cutoff at that redshift. Neither analytic parent distribution is expected to reproduce the lensed samples directly, because the probability that an ETG acts as a strong lens also depends on the lensing geometry and cross-section. This lensing weighting shifts both the full and observable lensed samples toward lower $z_d$ relative to the input ETG population, while the extended high-redshift tail reflects the broader $z_d\leq4.5$ deflector range adopted in this work. The additional observational selection produces only a modest change, primarily by slightly suppressing the high-redshift tail where lensing efficiency and source detectability are lower.

    Observational selection produces a clear bias toward lower source redshifts. The full lensed sample spans $4.5067$--$8.4932$ in $z_s$, with a median of $5.7485$, slightly higher than the input LRD median of $z_s=5.6513$. By contrast, the observable sample spans $4.5067$--$7.8176$, with a lower median of $5.3916$ (middle-right panel of Figure~\ref{fig:lensed_lrd}). This bias is driven mainly by the rapid decrease in the observed flux of LRDs toward higher redshift, which makes the fainter lensed images more likely to fall below the limiting magnitude. Thus, although high-redshift LRDs can still be strongly lensed in the ideal mock sample, a smaller fraction remain detectable after observational selection.

    Observational selection has little effect on the overall distribution of maximum image separations. In the full lensed sample, the maximum separation ranges from 0.0727 to 5.3047 arcsec, with a median of 1.1902 arcsec; in the observable sample, it ranges from 0.1493 to 4.9808 arcsec, with a median of 1.1966 arcsec (bottom-left panel of Figure~\ref{fig:lensed_lrd}). The lower bound increases because systems below the adopted PSF-resolution criterion are unresolved. The upper bound decreases mainly because some large-separation systems have higher source redshifts and therefore fainter apparent magnitudes, causing them to fall below the detection limit. 

    The bottom-right panel of Figure~\ref{fig:lensed_lrd} shows observable surface densities of $3.70\pm1.89~{\rm deg}^{-2}$ for doubles and $0.52\pm0.58~{\rm deg}^{-2}$ for quads. We estimate these values by grouping the simulated systems according to their {\tt square\_degree\_id} and counting the number of lenses in each of the 50 independent $1~{\rm deg}^{2}$ fields. The quoted uncertainties are the sample standard deviations of the field-by-field counts and therefore quantify the expected field-to-field variation. Scaling these surface densities to $50~{\rm deg}^{2}$ gives approximately 185 observable doubles and 26 observable quads. The shaded regions show the corresponding scatter in the expected number of systems over a survey area $A$, calculated as
    \[
    \sigma_N(A)=\sigma_{1,{\rm deg}^2}\sqrt{A},
    \]
    assuming that the $1~{\rm deg}^{2}$ fields are independent. The fractional field-to-field scatter is larger for quads because of their lower surface density.

    We quantify the significance of a null detection with a Poisson model based on the mock-predicted observable surface densities. For a survey area $A$, the expected number of detectable systems is $\mu=rA$, where $r$ is the relevant surface density, and the probability of detecting no systems is $P(N=0)=e^{-\mu}$.

    As an extended nominal reference, we adopt an approximate cumulative nominal JWST/NIRCam footprint of $1.19~{\rm deg}^{2}$. We first combine the de-duplicated nominal footprints of the six fields used above to derive the representative limiting magnitudes---CEERS, PRIMER-COSMOS, COSMOS-Web, PRIMER-UDS, JADES GOODS-S, and JADES GOODS-N---which together cover $2392~{\rm arcmin}^{2}$. We then add the nominal $1455~{\rm arcmin}^{2}$ POPPIES footprint \citep{Kartaltepe2024} and the $432~{\rm arcmin}^{2}$ of novel sky covered by PANORAMIC in at least six broad bands \citep{Williams2025}. Using the novel-sky area for PANORAMIC avoids double-counting pre-existing JWST coverage. Thus,
    \begin{equation}
    A_{\rm nom}
    =
    \frac{
    2392 + 1455 + 432
    }{3600}
    \simeq 1.19~{\rm deg}^{2}.
    \end{equation}

    Because POPPIES and PANORAMIC are pure-parallel programs assembled from heterogeneous pointings, neither program has a single uniformly defined limiting depth across its full footprint. When including their areas, we therefore assume that their effective limiting magnitudes are the same as the representative values derived above by area-weighting the six primary survey fields in flux space. Given this assumption, the differences in filter coverage among the constituent programs, and the possibility of smaller residual overlaps, the resulting $1.19~{\rm deg}^{2}$ should be interpreted as an approximate geometric footprint rather than a rigorously homogenized effective search area.

    At this extended nominal area, the model predicts null-detection probabilities of approximately $1.2\%$ for observable doubles, $53.9\%$ for observable quads, and $0.66\%$ for the total observable lensed-LRD sample. If we instead anchor the calculation to only the six de-duplicated fields used to determine the representative limiting magnitudes, with a combined area of $2392~{\rm arcmin}^{2}$ ($0.664~{\rm deg}^{2}$), the corresponding null-detection probabilities are $8.6\%$, $70.8\%$, and $6.1\%$, respectively. Comparing these values with the one-sided Gaussian tail probabilities corresponding to $3\sigma$ and $5\sigma$, we infer the survey areas required for a non-detection to reach these significance thresholds, as summarized in Table~\ref{tab:non_detection_area}.

\begin{table}
    \caption{Required cumulative survey area (in square degrees) for a non-detection to correspond to one-sided $3\sigma$ and $5\sigma$ tail probabilities, assuming the area-weighted $5\sigma$ limiting magnitudes given in Equation~\ref{eq:weighted_depth}.}
    \label{tab:non_detection_area}
    \begin{tabular}{lcc}
    \toprule
     & $3\sigma$ & $5\sigma$ \\
    \midrule
    Double & 1.786 & 4.072 \\
    Quad & 12.707 & 28.971 \\
    Total & 1.566 & 3.570 \\
    \bottomrule
    \end{tabular}
\end{table}

\section{Discussion}
\label{sec:diss}
\subsection{The Searching Strategy}
\label{sec:diss_search}

    In Section~\ref{sec:res}, we showed that there is a high probability that double-image systems are already present in existing survey data. The natural question is therefore how such systems can be identified. Searches for lensed LRDs may differ substantially from conventional searches for lensed AGNs, in which point-like quasar images can often be recognized directly around a foreground galaxy \citep{lemon2022,He2025aa}. By contrast, lensed LRD images are generally faint relative to the foreground deflector light and may therefore be easily overlooked.

\begin{figure*}
    \includegraphics[scale=0.42]{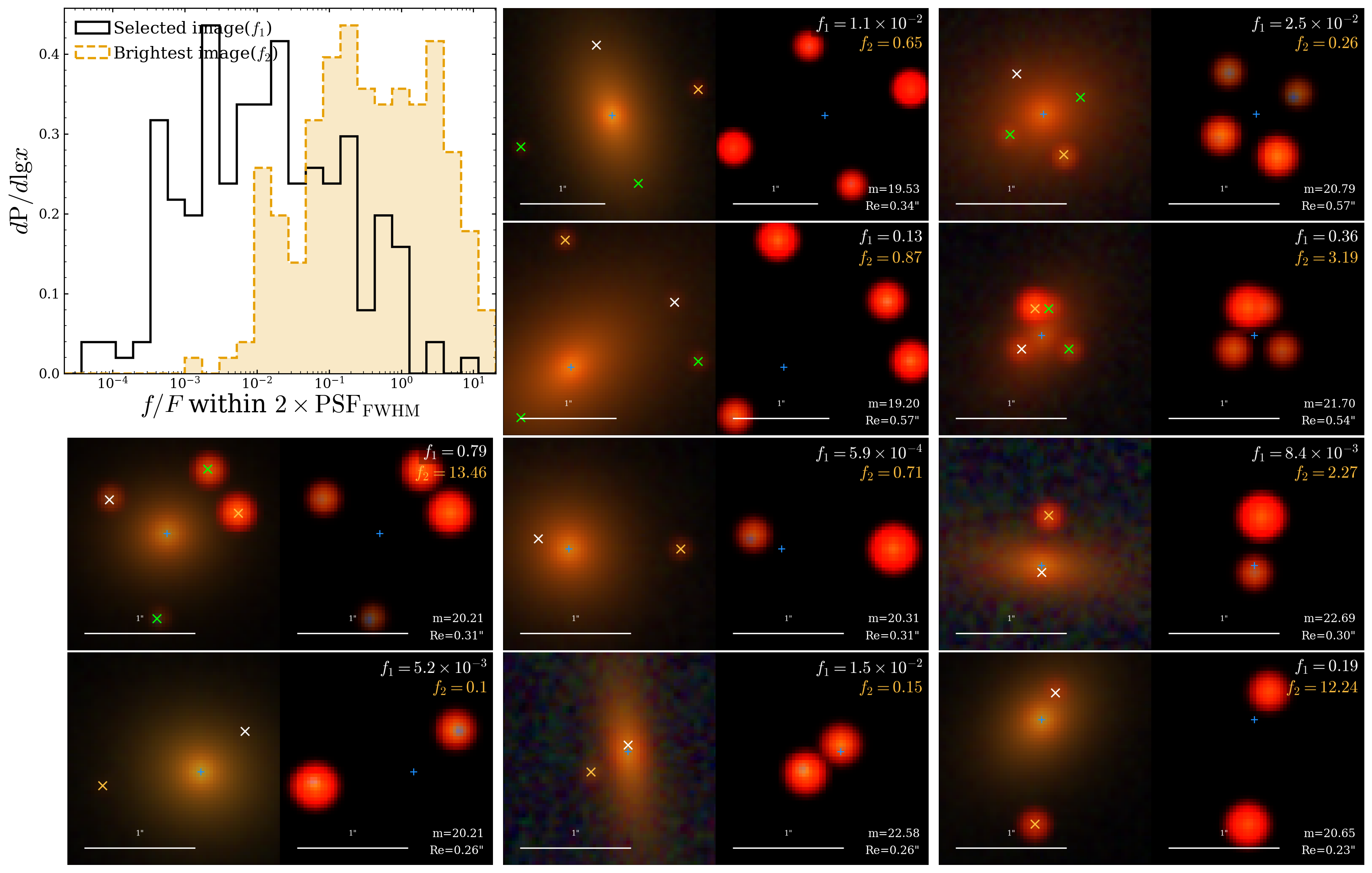}
    \caption{Foreground-light contamination in mock observable lensed LRDs. The histogram shows the distribution of the flux ratio $f/F$, measured within a circular aperture of diameter $2\times{\rm PSF}_{\rm FWHM}$, where $f$ is the flux of a lensed image and $F$ is the local flux contributed by the foreground lens. The solid histogram corresponds to the selected image, defined as the second-brightest image in doubles and the third-brightest image in quads, while the dashed histogram corresponds to the brightest lensed image. The image stamps show representative systems. In each pair, the left panel presents the mock observed image, including foreground-lens light and noise, whereas the right panel shows the corresponding noise-free, PSF-convolved lensed-source image with the foreground deflector omitted. The magnitude and effective radius of each deflector are listed in the lower-right corner of the corresponding right-hand panel. Blue crosses mark the centers of the foreground deflectors. Orange and white crosses indicate the brightest and selected lensed images, respectively, while the remaining lensed images are marked by green crosses. The definitions of $f_1$ and $f_2$ are given in Section~\ref{sec:diss_search}.}
    \label{fig:etg_contamination}
\end{figure*}

    This difficulty is illustrated by the color-composite images of mock lensed LRDs shown in Figure~\ref{fig:etg_contamination}. In most observable systems, the multiple images of the LRD are difficult to identify without first subtracting the foreground deflector light. To quantify this effect, we define two flux ratios between the lensed images and the local ETG emission, measured within circular apertures of diameter $2{\rm FWHM}_{\rm PSF}$. We denote by $f_1$ the flux ratio for the image used in our detectability criterion and by $f_2$ the corresponding ratio for the brightest lensed image in each system. All measurements are performed in the F444W band, in which LRDs are typically brightest.

    Several factors make foreground ETG light a major obstacle to searches for lensed LRDs. First, the images selected by our detectability criterion are typically much fainter than the local ETG emission, with
    \[
    f_1 = 0.015_{-0.014}^{+0.174},
    \]
    and can therefore be easily missed. Second, even the brightest lensed image often remains fainter than the foreground emission. The corresponding flux ratio is
    \[
    f_2 = 0.349_{-0.308}^{+2.755},
    \]
    although the distribution has an extended tail towards high flux ratios. Third, some multiple images are projected onto regions of particularly high ETG surface brightness, making their recovery highly sensitive to imperfect lens-light subtraction, PSF mismatch, and residual structures in the foreground-light model.

    Several strategies may help mitigate contamination from the deflector light. The most direct approach is to first identify ETGs in the survey field and then perform image modelling to separate the lensed-LRD emission from the foreground galaxy light. This should be feasible because ETGs generally have relatively regular morphologies, and their surface-brightness distributions can often be described using one or more S\'ersic components. A second possibility, inspired by color-based searches for strongly lensed supernovae \citep{Quimby2014,Magee2023}, is to exploit the strong color contrast between ETGs and LRDs. Subtracting appropriately scaled images obtained in bluer and redder bands may suppress the foreground ETG emission while preserving the unusually red lensed images. Finally, convolutional-neural-network-based lens finders may provide an effective complementary approach, as they can be trained to recognize faint and partially blended lensing features that are difficult to identify through visual inspection or simple catalog-level selection \citep{Wang2022,He2025apj}.

\subsection{Limitations and Future Prospects}

    Several limitations of the present analysis should be noted. First, the mock background population includes a simplified treatment of both the LRD luminosity function and spatial distribution. We adopt the best-fitting Schechter parameters reported by K24 and fix $(M_\ast,\phi_\ast,\alpha)$ throughout the simulation. The uncertainties in these parameters, particularly in the faint-end slope $\alpha$, could substantially affect the predicted surface density of lensed LRDs. A consistent propagation of these uncertainties requires their covariance matrix or joint posterior distribution because the three parameters are strongly correlated. As these quantities are currently unavailable, Table~\ref{tab:non_detection_area} provides the corresponding $3\sigma$ and $5\sigma$ estimates as practical reference values. We also distribute LRDs uniformly across the source plane. Emerging observations of dual LRD systems and LRDs in galaxy overdensities indicate enhanced close-pair incidence on kiloparsec scales \citep{Tanaka2024,Merida2025}, suggesting that intrinsic clustering may alter both the mean lensing abundance and its field-to-field variation. Measurements over substantially larger areas will enable a direct determination of the LRD luminosity function, its parameter covariance, and the clustering properties of the population.

    Second, our treatment of the observational selection function and survey geometry is simplified. We derive a combined $5\sigma$ limiting magnitude in each NIRCam band using an area-weighted average across the selected survey fields. This approximation captures the typical sensitivity of existing large-area data sets while averaging over field-to-field differences in depth, filter coverage, and angular resolution. Furthermore, based on an extrapolation from the design lifetime of JWST and the cumulative volume of data released to date, the final NIRCam area suitable for lensed-LRD searches is expected to be only $\sim10~{\rm deg}^{2}$. Our adopted $50~{\rm deg}^{2}$ area should therefore be interpreted solely as the Monte Carlo simulation domain rather than as a prediction of the eventual JWST survey footprint. All predicted abundances are normalized per square degree, and the larger simulated area is used only to reduce Monte Carlo sampling noise and obtain more stable estimates of the mean lensing abundance and field-to-field scatter of these rare systems. Nevertheless, we do not explicitly quantify how the finite size of the simulation domain affects the uncertainties in the inferred lensing surface densities and their field-to-field variance.

\section{Conclusion}
\label{sec:conclu}
    We have presented the first benchmark estimate of the expected abundance of galaxy-scale strongly lensed LRDs in JWST surveys by combining a literature-based LRD population model with a foreground ETG lens population. In our fiducial mock survey, the intrinsic lensing surface densities are $10.70\pm3.76~{\rm deg}^{-2}$ for doubles and $0.64\pm0.69~{\rm deg}^{-2}$ for quads. After applying a JWST-based observational transfer function that accounts for finite PSF resolution and limiting magnitude, these rates decrease to $3.70\pm1.89~{\rm deg}^{-2}$ and $0.52\pm0.58~{\rm deg}^{-2}$, respectively.

    The absence of securely confirmed systems is therefore substantially more constraining for doubles than for quads. Anchoring the calculation to the de-duplicated ($0.66~{\rm deg}^{2}$) footprint covered by  COSMOS-Web, PRIMER-UDS, PRIMER-COSMOS, CEERS, JADES GOODS-S, and JADES GOODS-N yields non-detection probabilities of $8.6\%$ for doubles and $70.8\%$ for quads. Using an extended nominal geometric footprint of $\sim1.19~{\rm deg}^{2}$, which additionally includes POPPIES and PANORAMIC and assumes the same representative limiting depths derived for the six primary survey fields, reduces these probabilities to $1.2\%$ and $53.9\%$, respectively. This marked contrast suggests that detectable lensed LRDs, particularly doubles, may already be present in existing JWST imaging but have so far been overlooked or misclassified.

    We further find that contamination from foreground ETG light may represent a major obstacle to detection, as individual lensed images are typically much fainter than the local surface brightness of the lens galaxy. We therefore propose several complementary search strategies to mitigate this problem, including accurate modelling and subtraction of the lens-galaxy light, difference imaging between blue and red bands, and CNN-based identification methods. With improved search techniques and the continued accumulation of JWST imaging data, we expect galaxy-scale strongly lensed LRDs to be identified in the near future, providing a powerful new probe of the physical nature of the LRD population.


\bibliography{sample701}{}
\bibliographystyle{aasjournalv7}



\end{document}